\documentclass[prb,twocolumn,superscriptaddress,citeautoscript,floatfix]{revtex4-1}

\usepackage[bookmarks=false,linkcolor=blue,urlcolor=blue,colorlinks,citecolor=blue]{hyperref}
\usepackage{graphicx,soul}
\usepackage{color}
\usepackage{amsmath}
\usepackage{amsfonts}
\usepackage{amssymb}
\usepackage{bm}

\begin{document}
	
\title{Topological polaritons from photonic Dirac cones coupled to excitons \\in a magnetic field}

\author{Kexin Yi and Torsten Karzig}
\affiliation{Institute for Quantum Information and Matter, Caltech, Pasadena, California 91125, USA}

\begin{abstract}
We introduce an alternative scheme for creating  topological polaritons (topolaritons) by exploiting the presence of photonic Dirac cones in photonic crystals with triangular lattice symmetry. As recently proposed, topolariton states can emerge from a coupling between photons and excitons combined with a periodic exciton potential and a magnetic field to open up a topological gap.  We show that in photonic crystals the opening of the gap can be substantially simplified close to photonic Dirac points. Coupling to Zeeman-split excitons breaks time reversal symmetry and allows to gap out the Dirac cones in a nontrival way, leading to a topological gap similar to the strength of the periodic exciton potential. Compared to the original topolariton proposal [Karzig {\em et al}, PRX {\bf 5}, 031001 (2015)], this scheme significantly increases the size of the topological gap over a wide range of parameters. Moreover, the gap opening mechanism highlights an interesting connection between topolaritons and the Haldane and Raghu scheme [Haldane and Raghu, PRL {\bf 100}, 013904 (2008)] to create topological photons in magneto-optically active materials.
\end{abstract}
\maketitle

\section{Introduction}
Topological electronic systems play a crucial role in modern condensed matter physics. Recently, it became possible to extend the idea of topologically non-trivial bandstructures to photonic and other bosonic systems, which led to the field of topological photonics \cite{lu_topological_2014}. This extension is far from trivial since Kramer's theorem (which protects electronic topological insulators) cannot be applied to bosons. Moreover, photons lack a natural equivalent of the orbital magnetic field that is used to create a quantum Hall effect, i.e. time-reversal-symmetry broken topological states. Proposals for topological photons therefore rely on the geometric Berry curvature which can emulate magnetic field effects. In the electronic context this gives rise to the quantum anomalous Hall effect \cite{haldane_model_1988,yu_quantized_2010}, which has been observed recently in ferromagnetically doped topological insulators \cite{chang_experimental_2013}.

The idea to create quantum-(anomalous)-Hall-like states of photons was first discussed in 2008 by Haldane and Raghu.\cite{haldane_possible_2008,raghu_analogs_2008} Their proposal relies on photonic Dirac cones that appear in specific (e.g., triangular or hexagonal) photonic crystals. Breaking time reversal symmetry for photons introduces a topologically non-trivial gap resulting in bands characterized by non-zero Chern numbers and chiral photonic edge modes. The necessary time reversal symmetry breaking can be provided by magneto-optically active materials which allows for a straightforward experimental realization of topological electromagnetic modes in the microwave regime \cite{wang_observation_2009,poo_experimental_2011}. At optical frequencies, however, the corresponding magneto-optical effects are too weak \cite{landau_electrodynamics_1960}. Concepts that were recently developed to avoid this difficulty include the use of (quasi) time periodic systems \cite{fang_realizing_2012,rechtsman_photonic_2013} or suitably coupled optical cavities or resonators \cite{,cho_fractional_2008,koch_time-reversal-symmetry_2010,umucalilar_artificial_2011,yannopapas_photonic_2011,hafezi_robust_2011,hafezi_imaging_2013,ningyuan_time-_2015}. Another promising direction are hybrid systems where photons interact with mechanical\cite{peano_topological_2015} or excitonic \cite{karzig_topological_2015,nalitov_polariton_2015,bardyn_topological_2015,bardyn_chiral_2015,yuen-zhou_plexcitons_2015} degrees of freedom. In particular it is possible to break time reversal symmetry  by coupling photons to Zeeman-split excitons\cite{karzig_topological_2015,nalitov_polariton_2015,bardyn_topological_2015}. This can give rise to a winding exciton-photon coupling resulting in non-trivial Chern numbers of the hybridized (polariton) bands. 

Here, we show that this winding coupling can be used to (non-trivially) gap out photonic Dirac cones, giving rise to an alternative way of creating topological photonic states. Interestingly, this scheme allows to create larger topological gaps in comparison to the original proposal for topological polaritons \cite{karzig_topological_2015} in which photonic Dirac cones were not taken into account. Moreover, it is also conceptually appealing since it allows to draw parallels between topological polaritons and Haldane and Raghu's proposal for topological photons \cite{haldane_possible_2008,raghu_analogs_2008}.

\section{Gapping out Dirac cones}
\label{sec:Dirac}

The topological properties of quantum-anomalous Hall states are characterized by the Chern numbers of the bandstructure. The latter are defined by the Berry flux of each band  (divided by $2\pi$), which always takes integer values in weakly interacting systems. The reason why it is particularly promising to search for topological states close to Dirac points of the bandstructure, is the well known fact that each (gapped) Dirac cone contributes a Berry flux of $\pm\pi$. The way the Dirac cones are gapped out then decides whether the Berry fluxes of the even number of Dirac cones in the bandstructure add up to finite or vanishing Chern numbers \cite{haldane_model_1988}.

More specifically, let us follow the example of Haldane and Raghu \cite{haldane_possible_2008,raghu_analogs_2008} of a triangular photonic crystal. With a suitable form of the photonic crystal, the hexagonal photonic Brillouin zone has two isolated Dirac points of transverse electric (TE) character (transverse magnetic (TM) modes are gapped at this frequency) at momenta $K$ and $-K$. The twofold degeneracy at the Dirac points is protected by a combination of time-reversal and inversion symmetry. Since the Berry curvature is an even(odd) function of momentum when inversion(time-reversal) symmetry is present, it vanishes when both symmetries are combined. The latter allows to define a basis where the eigenfunctions of Maxwell's equations can be chosen entirely real, leading to a vanishing Berry connection (up to possible singularities at degeneracy points) as realized by gapless Dirac cones.
	
\begin{figure}
	\includegraphics[scale=0.08]{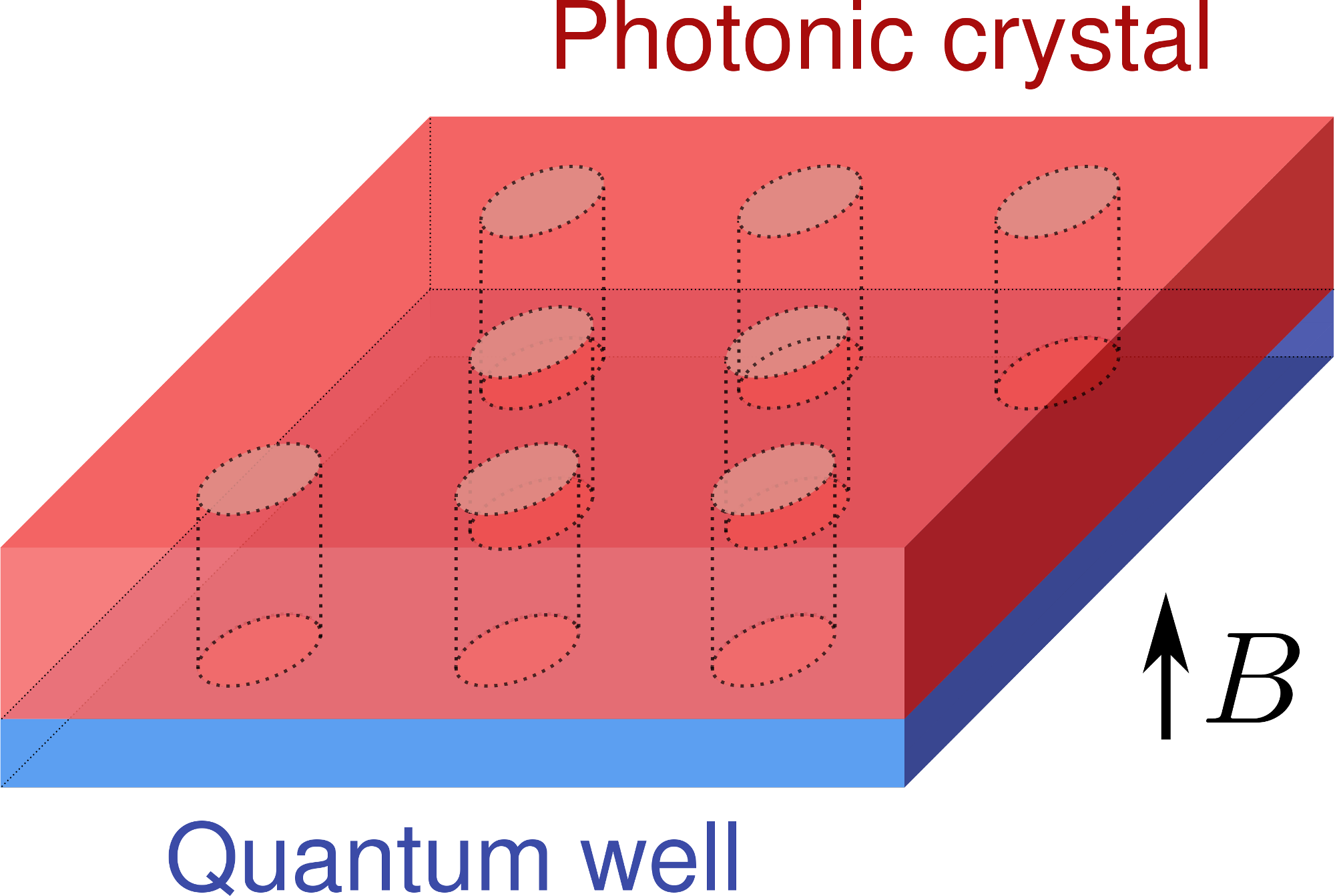}
	\caption{Schematic view of the setup. A triangular photonic crystal (red) is coupled to an exciton carrying quantum well (blue). Time reversal symmetry is broken due to a Zeeman splitting of the excitons in an external magnetic field.}
	\label{fig:setup}	
\end{figure}

Breaking time-reversal symmetry will in general gap out the Dirac cones and lead to non-trivial Chern numbers (while breaking inversion symmetry only leads to trivial gaps). In the case of the Haldane and Raghu proposal, time reversal is broken directly on the photon level due to magneto optical effects. Here we propose to break time-reversal symmetry by a resonant coupling of the photons to Zeeman-split excitons (for a more detailed comparison of the two approaches see Sec.~\ref{sec:comparison}). This can be accomplished by introducing exciton-carrying semiconductor quantum wells inside or close to the cavity that confines the photons to a two dimensional photonic crystal (see Fig.~\ref{fig:setup}). An applied magnetic field, then splits the bright excitons with total angular momentum projection $J_z=\pm1$ in energy. For simplicity we focus on a single species of excitons (with, say, $J_z=+1$), which is justified for large Zeeman splittings. \footnote{The results remain essentially unchanged when taking into account both excitons as long as the Zeeman splitting is larger than the topological gap of the single-exciton case (see, e.g., Ref.~\onlinecite{karzig_topological_2015})} 

Angular momentum conservation in the $z$-direction dictates that the coupling of the ($J_z=+1$) exciton to the linearly polarized ($J_z=0$) TE-photon exhibits a winding phase,
\begin{equation}
  H_{\rm X-P}=g_{q} \mathrm{e}^{\mathrm{i} \theta_\mathbf{q}}\hat{a}_\mathbf{q}^\dagger \hat{b}_\mathbf{q}+{\rm H.c.}\,
  \label{eq:winding-coupling}
\end{equation}
where $\hat{a}^\dagger_\mathbf{q}$($\hat{b}^\dagger_\mathbf{q}$) creates a photon(exciton) with momentum $\textbf{q}=(q_x,q_y)$, $g_q$ is the absolute value of the exciton-photon coupling, and $\theta_\mathbf{q}$ denotes the angle of the momentum vector relative to the $x$-axis.

\begin{figure}
	\includegraphics[scale=0.168]{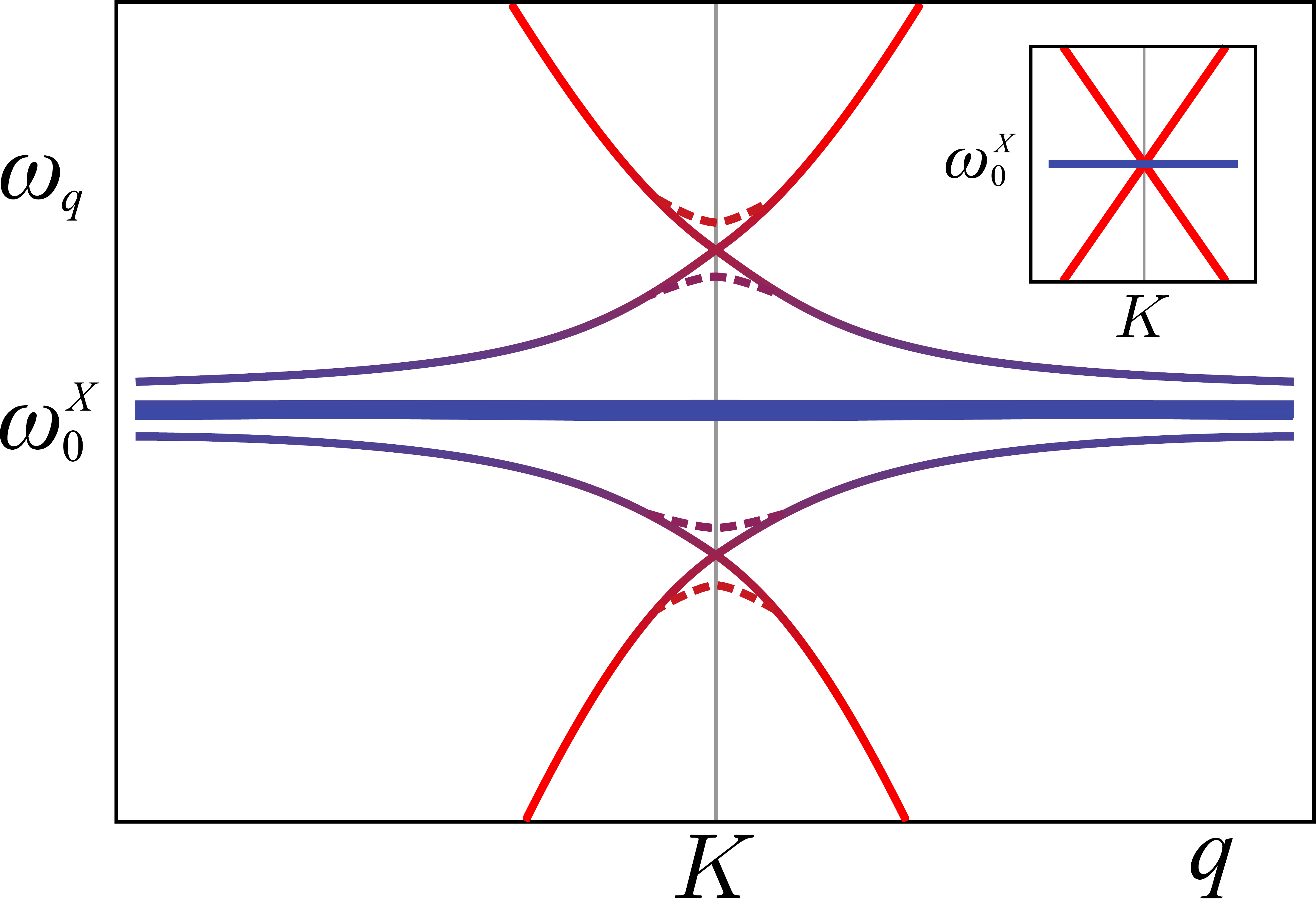}
	\caption{Schematic dispersion close to the photonic Dirac cone at $q=K$ without (inset) and with a finite exciton-photon coupling. The inset shows the position (here on resonance, $\delta=0$) of the bare exciton dispersion (blue) relative to the photonic Dirac cone (red). The main panel shows the splitting of the photonic Dirac cone into two polaritonic Dirac cones. The latter are gaped out when the exciton-photon coupling has a winding and an additional periodic exciton potential removes accidental degeneracies (dashed lines). The color coding indicates the strength of the excitonic (blue) and photonic (red) component in the polariton wavefunction.}
	\label{fig:Dirac_point}	
\end{figure}

We now consider the case where the excitons are resonant with the photonic Dirac cone. A finite exciton-photon coupling then splits the photonic Dirac cone into two polaritonic ones (see Fig.~\ref{fig:Dirac_point}). A winding (rather than constant) coupling then allows to gap out these Dirac cones because of the explicitly broken time-reversal symmetry. Although the Dirac cones are no longer symmetry-protected it is still possible to have accidental degeneracies. To avoid the latter we use an additional periodic exciton potential which leads to well-defined topological gaps with corresponding chiral edge modes.

\subsection{Theoretical estimate for the topological gap}
\label{sec:theo_estimate}

\begin{figure}
	\includegraphics[scale=0.8]{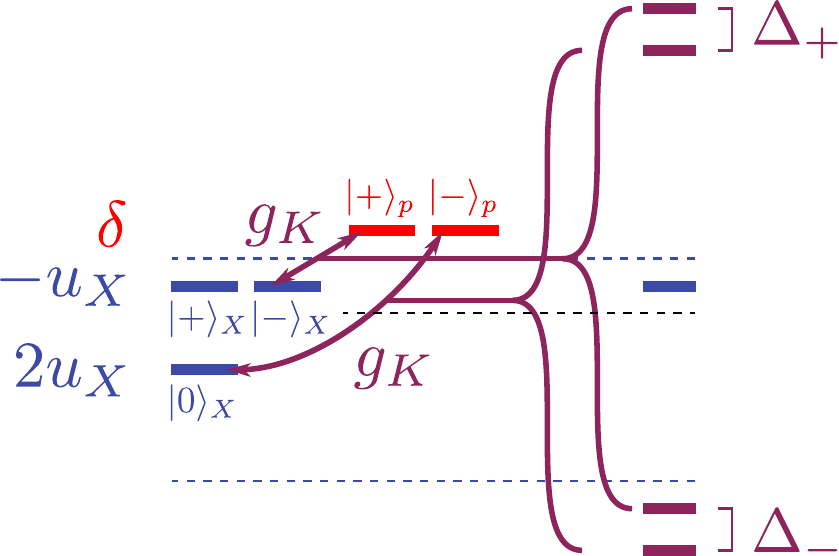}
	\caption{Schematic energy spectrum close to the photonic Dirac point with (right) and without (left) the presence of an exciton-photon coupling $g_K$.  All energies are measured relative to the free exciton energy $\omega_0^X$ indicated by a black dashed line. We choose a periodic exciton potential strength $u_X<0$ and neglect the higher energy photonic mode $|0\rangle_P$ ($u_P\gg g$). The dashed blue lines at $-2u_X$ and $6u_X$ indicate the energy range of the large-momentum excitons corresponding to the maximum and minimum of the excitonic potential defined in Eq.~(\ref{eq:potential}).}
	\label{fig:splitting}
\end{figure}

Here we use the example of a triangular lattice to show the effect of a winding coupling on photonic Dirac cones and discuss the possible accidental degeneracy that remains in the absence of a periodic exciton potential. To simplify the discussion of the emerging gap we focus on the states at one of the Dirac points of the Brillouin zone ($\mathbf{q}=K$). For a vanishing strength of the periodic photonic potential the photonic dispersions originating from the centers of three adjacent (hexagonal) Brillouin zones (BZ) are (three-fold) degenerate at the $K$ point with energy $\omega^{P}_K$, where $\omega^{P}_\mathbf{q}$ is the (free) dispersion of the cavity photons. A finite periodic potential then mixes these three states (corresponding to momenta $q_1=|K|(1,0), q_2=|K|(-1,\sqrt{3})/2,\ \mathrm{and}\ q_3=|K|(-1,-\sqrt{3})/2$) with equal strength $u_P$, leading to a symmetric superposition $|0\rangle_P=(1,1,1)_P$ which is split-off from two winding superpositions $|+\rangle_P=(1,\mathrm{e}^{\mathrm{i}2\pi/3},\mathrm{e}^{-\mathrm{i}2\pi/3})_P$ and $|-\rangle_P=(1,\mathrm{e}^{-\mathrm{i}2\pi/3},\mathrm{e}^{\mathrm{i}2\pi/3})_P$ by an energy $3u_P$. The latter are the two degenerate states at the Dirac point. The same arguments apply for the excitonic states  (with the label $P$ replaced by $X$). In the following we measure all energies relative to the free exciton energy $\omega_\textbf{K}^X\approx \omega_0^X$ and denote the energy detuning of the photonic Dirac point from the excitons as $\delta=E_{|+\rangle_P}$ (see Fig.~\ref{fig:splitting}).

A constant (non-winding) exciton-photon coupling mixes states with the same angular momentum, i.e.,  $|i\rangle_P$ with $|i\rangle_X$ ($i=-1,0,+1$). Since this preserves the symmetry between $|+\rangle$ and $|-\rangle$ subspaces there are remaining degeneracies that give rise to polaritonic Dirac points (see Fig.~\ref{fig:Dirac_point}) with energies $\pm g_K$ for $g_K\gg \delta,u_X$. This symmetry is broken by a winding coupling [cf. Eq.~\eqref{eq:winding-coupling}] which mixes $|i\rangle_P$ with $|(i+1)\,{\rm mod}\,3\rangle_X$ (see Fig.~\ref{fig:splitting}). A finite periodic exciton potential will split the $|0\rangle_X$ and $|-\rangle_X$ states, which allows to harness the time-reversal symmetry breaking by splitting the $(|+\rangle_P,|-\rangle_X)$- and $(|-\rangle_P,|0\rangle_X)$-polaritons. The interplay of the winding coupling and the periodic exciton potential therefore acts as a time-reversal-broken perturbation that gaps out the polaritonic Dirac cones, which is an explicit example of the mechanism of Haldane and Raghu\cite{haldane_possible_2008} working at optical frequencies. We denote the corresponding gaps above ($+$) and below ($-$) the exciton resonance $\omega_0^X$ with $\Delta_\pm$. In the limit of small $|u_X|\ll g_K$ we find
\begin{equation}
  \Delta_\pm=\frac{3}{2}|u_X| \left(1+ \frac{2\delta \pm u_X}{2\sqrt{\delta^2+4g_K^2}} \right)\,.
  \label{eq:gap}
\end{equation}
Typical sizes for the topological gap are therefore of the order of $3|u_X|/2$. When focusing on the lowest energy gap ($-$) it is optimal to choose $u_X<0$ (meaning that the periodic exciton and photon potentials have opposite signs).

Interestingly, the largest gaps $\Delta_\pm=3|u_X|$ can be reached in the strongly off-resonant regime $\delta\gg g_K$. Note, however, that in the derivation of Eq.~\eqref{eq:gap} we made the simplifying assumption to only focus on states from the three BZs adjacent to the $K$ point. This is only well justified for photons, since their larger momentum states of higher BZs are energetically well-separated from the Dirac point. The exciton dispersion, on the other hand, is almost flat (on photonic velocity scales) which makes the large- and small-momentum excitons having essentially the same energy. Since the exciton-photon coupling, however, is momentum conserving, large momentum excitons only couple to strongly off-resonant photon states which leads to an accumulation of exciton states (backfolded from higher BZs) around $\omega_0^X$. The range of these (bare) exciton states is indicated as a blue band in Figs.~\ref{fig:Dirac_point} and \ref{fig:splitting}.  A requirement for the validity of Eq.~\eqref{eq:gap} is therefore that the topological gap is well-separated from the bare exciton states, which sets a lower limit for the strength of the exciton-photon coupling. For the lower gap \footnote{When taking into account an infinite number of BZs the finite (positive) exciton mass would make the excitons fill up the entire region $\omega_q\geq\omega_0^X$ including the topological gap at the upper polaritonic Dirac point. Although the corresponding edge modes would be largely decoupled from these large momentum excitons for not too sharp edges, only the lower topological gap is truly protected.} and in resonance ($\delta=0$) this yields the condition $g_K>7|u_X|$ ($u_X<0$) for reaching $\Delta_-\approx 3/2|u_X|$. Although the strongly off-resonant regime can in principle lead to larger gaps ($\Delta_-\approx 3 |u_X|$), reaching this value requires larger $g_K>\sqrt{7\delta |u_X|}$ in order to avoid the large-momentum exciton states despite the less-effective off-resonant exciton-photon coupling. It should also be noted that it might not always be possible to reach the strongly off-resonant regime because the maximal values of $\delta$ are naturally cut-off by the bandwidth of the Dirac cone (i.e., when the excitons become resonant with other photonic bands).

\section{Numerical results}

\subsection{Tight binding model and chiral edge modes}	
	
To demonstrate the topological nature of the gapped Dirac cones in this scheme, we perform numerical calculations based on a triangular lattice model (with lattice constant $a = 1$). When implementing a finite system size we expect chiral polaritonic edge modes inside the topological gaps. As discussed above, in order to open a topological gap, periodic potentials for both photons $V_P$ and excitons $V_X$ must be present. The Hamiltonian of the lattice model is thus given by \footnote{Note that the lattice is merely a discretization of space independent of the periodic potentials.}
\begin{align}
\begin{split}
H = & \sum \limits_{\mathbf{q}} [\omega_{\mathbf{q}}^p \hat{a}_{\mathbf{q}}^\dagger \hat{a}_{\mathbf{q}} + \omega_{\mathbf{q}}^X \hat{b}_{\mathbf{q}}^\dagger \hat{b}_{\mathbf{q}} + (g_{q} \mathrm{e}^{\mathrm{i} \theta_\mathbf{q}} \hat{a}_{\mathbf{q}}^\dagger \hat{b}_{\mathbf{q}} + {\rm H.c.})] \\ 
&+ \sum \limits_{\bf{r}} [V_P({\bf{r}}) \hat{a}_{\bf{r}}^\dagger \hat{a}_{\bf{r}} + V_X({\bf{r}}) \hat{b}_{\bf{r}}^\dagger \hat{b}_{\bf{r}}]
\end{split}
\label{eq:H}
\end{align}
where $\hat{a}^{\dagger}_{\mathbf{q}}$, $\hat{b}^{\dagger}_{\mathbf{q}}$ are the creation operators of photons and excitons, with respective dispersion relations $\omega_{\mathbf{q}}^p$, $\omega_{\mathbf{q}}^X$ ($\hat{a}^\dagger_\mathbf{r}$ and $\hat{b}^\dagger_\mathbf{r}$ are the corresponding Fourier transformed operators).

The exciton-photon coupling constant has a winding structure $g_{\mathbf{q}} \propto (q_x + \mathrm{i} q_y)$ at low momenta. By setting the coupling strength to $g_K$ at the Dirac points, the lattice version (respecting the hexagonal symmetry of the BZ) of the coupling takes the form,
\begin{equation}
g_{q} \mathrm{e}^{\mathrm{i} \theta_\mathbf{q}} = \frac{g_K}{\sqrt{3}} \Big(\sin \tfrac{q_x}{2} \cos \tfrac{\sqrt{3}q_y}{2} + \sin q_x + \mathrm{i} \sqrt{3} \cos \tfrac{q_x}{2} \sin \tfrac{\sqrt{3}q_y}{2} \Big)\,.
\end{equation}
The photon and exciton dispersions $\omega_{\mathbf{q}}^p$, $\omega_{\mathbf{q}}^X$ also have the same lattice symmetry, given by:
\begin{align}
\omega_{\mathbf{q}}^P &= \frac{4}{3m_{P}} \big(3-\cos{\tfrac{q_x}{2}}\cos{\tfrac{\sqrt{3}q_y}{2}}-2\cos{q_x} \big) \\
\omega_{\mathbf{q}}^X &= \frac{4}{3m_{X}} \Big(3-\cos{\tfrac{q_x}{2}}\cos{\tfrac{\sqrt{3}q_y}{2}} -2\cos{q_x}\Big) + \omega_{0}^X 
\end{align}
Here $m_{P}$, $m_{X}$ are the effective photon and exciton masses with $m_{X}\gg m_{P}$. The overall shift $\omega_{0}^X$ is the exciton energy required for the creation of a bounded electron-hole pair. The simplest implementation of the periodic exciton and photon potentials is to introduce a triangular superlattice with lattice constant $a_s=2$ (where the superlattice sites are shifted by an energy $6u_{p,X}$ while indermediate sites are shifted by $-2u_{p,X}$). The analytical form of these periodic exciton and photon potentials is given by
\begin{align}
V_{p,X}(\mathbf{r}) &= 2 u_{p,X} f(\mathbf{r}) \label{eq:potential} \\ 
f(\mathbf{r}) & = \cos\tfrac{2 \pi y}{\sqrt{3}} + \cos\big[\big(x+\tfrac{y}{\sqrt{3}}\big)\pi \big] + \cos\big[\big(x-\tfrac{y}{\sqrt{3}}\big)\pi\big]\,,
\end{align}

\begin{figure}
	\includegraphics[scale=0.19]{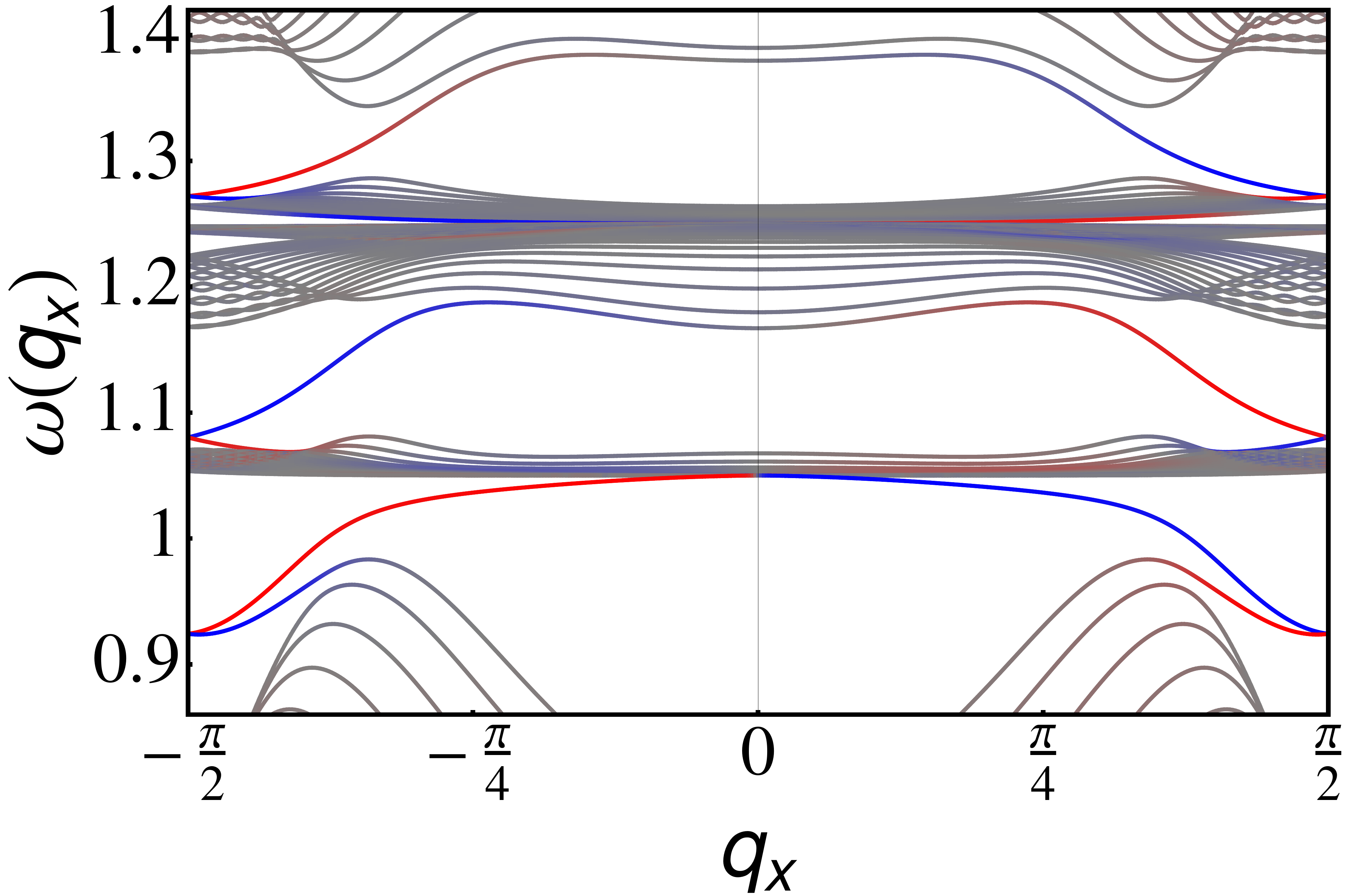}
	\caption{Spectrum of the tight binding model on a cylinder geometry. The gray lines denote bulk bands while the coloring indicates localization at one of the edges of the system with red(blue) coloring corresponding to the edge at $y=0(L)$. Used parameters: $m_P = 1$, $m_X = 10000$, $\omega_0^X = 1.2$, $g_0 = 0.1$, $u_P = 0.25$, $u_X = -0.025$, $L=35$.}
	\label{fig:Lattice_Model_Edgemodes}	
\end{figure}

To show the topological edge modes of the tight binding Hamiltonian~\eqref{eq:H} we put the system on a cylinder (Dirichlet boundary condition on the $y$ direction and periodic boundary condition in $x$ direction). The resulting spectrum is depicted in Fig.~\ref{fig:Lattice_Model_Edgemodes}. The bulk bands (shown in gray) show well-defined gaps due to the broken time reversal symmetry, while the colored in-gap states (red and blue corresponding to the edge at $y=0$ and $y=L$, respectively) are the edge modes resulting from the non-trivial topology. The expected topological gaps of Fig.~\ref{fig:Dirac_point} correspond to the lowest and highest energy gap in the depicted tight-binding spectrum. The chirality of the edge modes (in this case right handed) is determined by the sign of the winding coupling in Eq.~\ref{eq:winding-coupling}. As expected from Eq.~\ref{eq:gap} we indeed find that the topological gap of the lower polariton is optimized by choosing the sign of $u_X$ opposite to $u_P$.

The additional topological gap at energy $\approx\omega_0^X$ is an artifact of the lattice discretization. As mentioned in Sec.~\ref{sec:theo_estimate}, in the (continuum) limit $a\rightarrow 0$, the energy range $\{\omega_0^X-2u_X,\omega_0^X+6u_X\}$ (corresponding to energies $\{1.25,1.05\}$ in Fig.~\ref{fig:Lattice_Model_Edgemodes}) will be densely populated by a large number of (bare) excitonic bands such that only the upper and lower gaps remain.

\begin{figure}
	\includegraphics[scale=0.2]{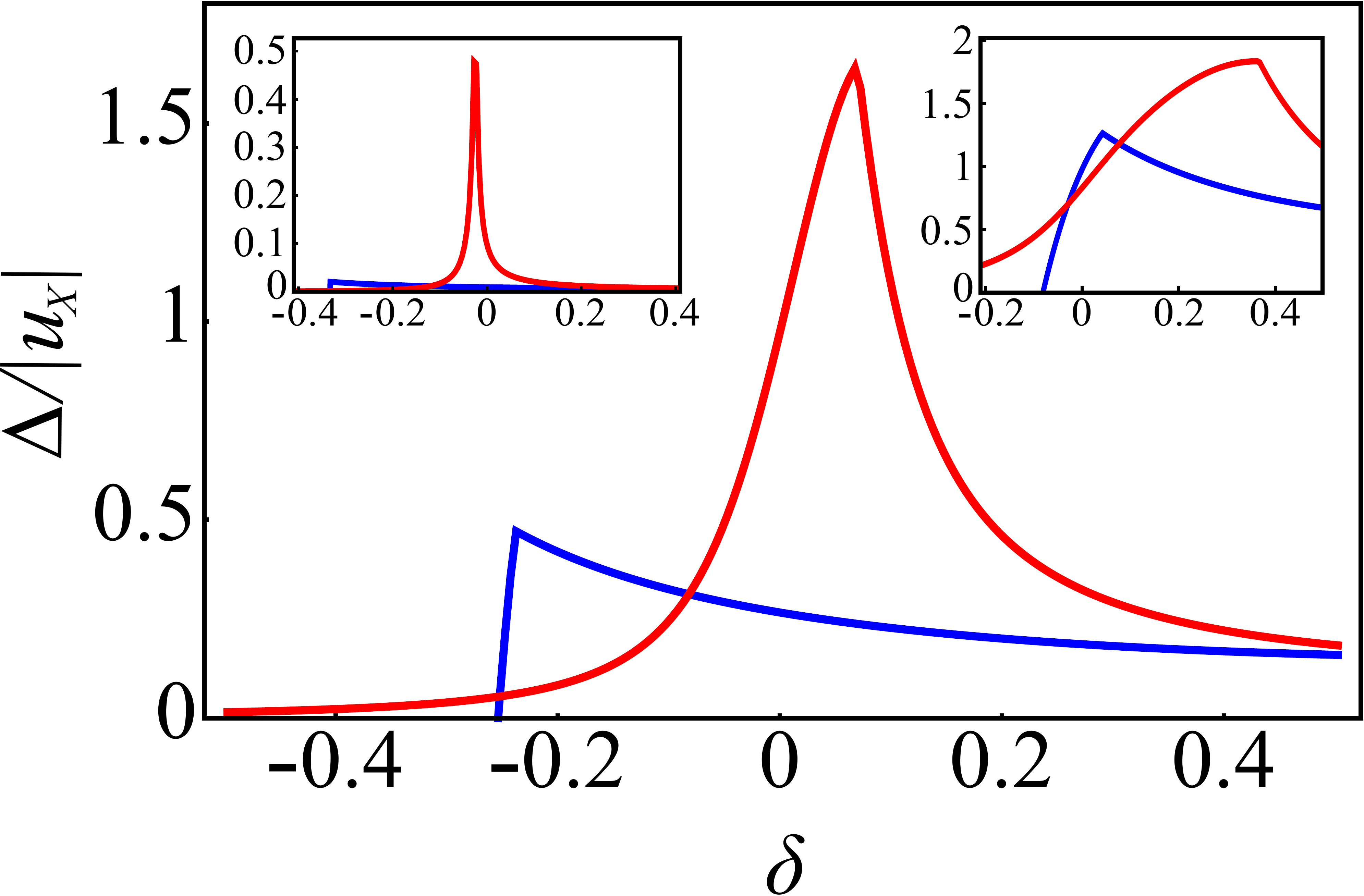}
	\caption{Size of the (lowest) topological gap for different parameters. The red curve shows the gap of the scheme presented in this paper ($u_P = 1.0$) with the main panel corresponding to $g_K = 0.05$ while the left and right insets show $g_K = 0.01$ and $g_K = 0.1$, respectively. Other parameters are given by $m_X = 5000$, and  $u_X = -0.005$. For small $g_K$ the largest gaps can be obtained when the excitons are tuned close to the photonic Dirac point ($\delta\approx 0$). The blue curve shows the original scheme\cite{karzig_topological_2015} (no  isolated Dirac points) for comparison using the same parameters except for $u_P=0$. As discussed in the main text the gap opening scheme of this paper allows for much larger gaps when the exciton-photon coupling is not too large.}
	\label{fig:Gap_Size_Comparison}	
\end{figure}

\begin{figure*}
	\includegraphics[width=2\columnwidth]{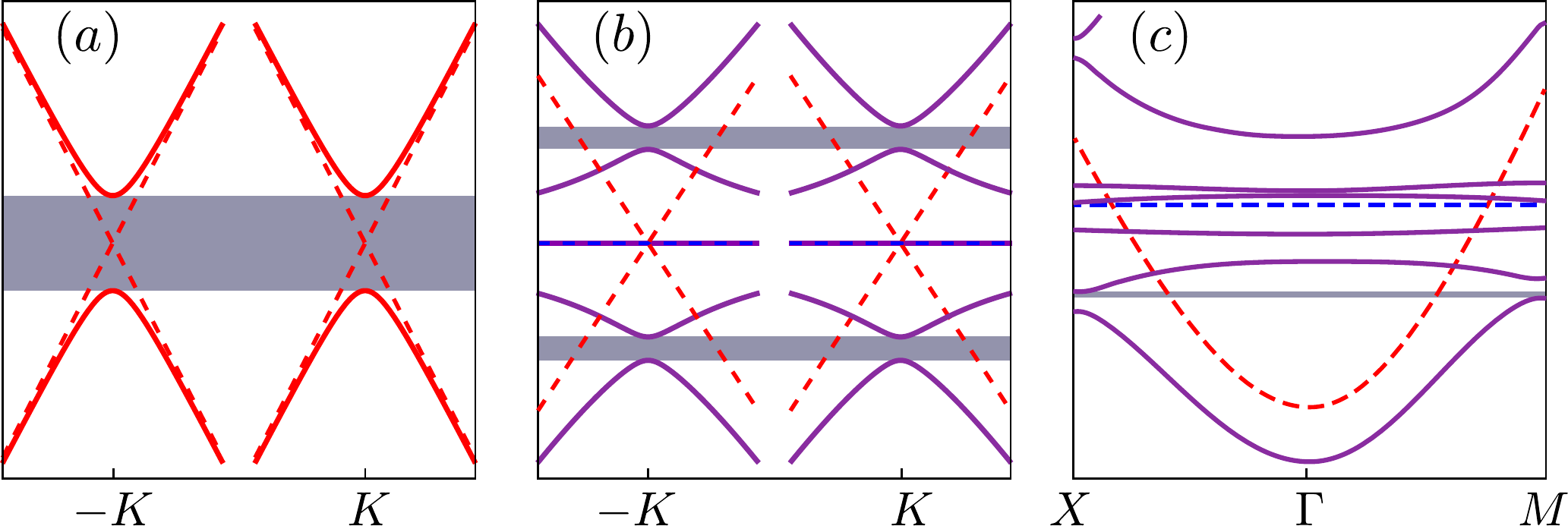}
	\caption{Comparison of (schematic) bandstructures of different topological photon/polariton proposals. The corresponding topological gaps are marked by gray regions. (a) Haldane and Raghu\cite{haldane_possible_2008,raghu_analogs_2008} proposal. Photonic Dirac cones of TE character (dashed) are gapped out by a finite magneto optical effect, thus creating bands with non-trivial topology (solid red). (b) The proposal presented in this paper (see Fig.~\ref{fig:Dirac_point}). TE photonic Dirac cones (dashed red) are resonantly coupled to a single (say, $J_z=+1$) exciton species (dashed blue). The hybridized modes (solid purple) exhibit gapped-out Dirac cones which lead to topological bands similar to (a). (c) Original topological polariton proposal \cite{karzig_topological_2015}. The non-trivial topology of the polariton modes (solid purple) emerges from a winding coupling of TE photons (dashed red) and a single (say, $J_z=+1$) exciton species (dashed blue) without the need of any Dirac cones. Since the band gaps that form at the edge of the (here quadratic) Brillouin zone (defined by the periodic exciton potential) are shifted between the $M=(\pi/a_s,\pi/a_s)$ and $X=(\pi/a_s,0)$ point, the size of the overall topological gap is reduced.}
	\label{fig:comparison}
\end{figure*}

\subsection{Size of the topological gap}

Although the lattice model gives a straightforward way to show the non-trivial topology of the Hamiltonian \eqref{eq:H} for quantitative estimates of the gap size we use a continuum model. This allows to implement the photon and exciton dispersions directly as $\omega_{\mathbf{q}}^p = c\sqrt{q^2+q_d^2}$ and $\omega_{\mathbf{q}}^X = q^2/2m_X + \omega_{0}^X $, where $q_d=\pi/d$ is given in terms of the thickness $d$ of the photonic cavity (we set $q_d=0$ for concreteness\footnote{The main effect of a finite $q_d$ can be described my renormalizing $c\rightarrow c_\mathrm{eff}=K/\sqrt{K^2+q_d^2}$ and an energy shift $\omega_0^X \rightarrow \omega_0^X-c_\mathrm{eff} q_d^2/K$}). The winding coupling and periodic potentials are given by Eq.~\eqref{eq:winding-coupling} with $g_q=g_K$ and Eq.~\eqref{eq:potential}, respectively. 
To compare the gap opening scheme presented here with the original proposal \cite{karzig_topological_2015}, where an explicit periodic photon potential and photonic Dirac cones were absent, we plot the (lowest) topological gap of both schemes for different values of $g_K$ and $\delta$ in Fig.~\ref{fig:Gap_Size_Comparison}.

Our results show that opening a gap close to the Dirac point will in general lead to larger gaps than the original scheme for a wide range of parameters (for an intuitive explanation see Sec.~\ref{sec:comparison}). The difference is especially pronounced for smaller exciton-photon couplings $g_K$. The latter observation follows directly from the estimate of Sec.~\ref{sec:theo_estimate}, where we determined that even  for relatively small exciton-photon couplings $g_K \gtrsim |u_X|$ we expect gap sizes of the order of $|u_X|$, when the excitons are in resonant with the photonic Dirac cone. In the absence of Dirac points reaching a global gap of similar size is much more demanding, requiring $g\gtrsim \sqrt{|u_X|\delta_P}$, where $\delta_P\sim c/a_s$ is of the order of the photonic band width, i.e. typically $\sim 1\mathrm{eV}$, which is much larger than typical $|u_X|$. For more moderate $g$, the gap in the original proposal is suppressed as  $g^2/\delta_P$  \cite{karzig_topological_2015}.

Also note that the results of Fig.~\ref{fig:Gap_Size_Comparison} agree very well with our previous estimate. For small $g_K$ we recover that the largest gaps $\approx 3/2|u_X|$ are obtained when the excitons are resonant with the photonic Dirac cone ($\delta=0$). For larger exciton photon-couplings increasing the detuning $\delta$ leads to increasing gaps approaching $\Delta_-=3|u_X|$ (cut-off when $\delta$ exceeds values $\sim g_K^2/7|u_X|$).

\section{Comparison of topological photon/polariton proposals}\label{sec:comparison}

We now compare the different existing topological polariton proposals\cite{karzig_topological_2015,nalitov_polariton_2015,bardyn_topological_2015} with the one discussed in this paper and the Haldane and Raghu proposal\cite{haldane_possible_2008,raghu_analogs_2008} for topological photons. Figure~\ref{fig:comparison}(a) shows the bandstructure of the Haldane and Raghu proposal, with (solid lines) and without (dashed lines) a magneto-optical coupling. As mentioned in Sec.~\ref{sec:Dirac}, breaking time reversal symmetry via the magneto-optical effect will gap out the Dirac cones at $K$ and $-K$. Each Dirac cone then contributes a Berry flux of $\pm\pi$ which add up to non-trivial Chern numbers $\pm1$ in the gapped out bands. In a similar way, the proposal presented here creates a non-trivial topology by gapping out the polaritonic Dirac cones that are created when hybridizing an exciton with the photonic Dirac cone [see Fig.~\ref{fig:comparison}(b)]. The corresponding time reversal symmetry breaking is provided by the excitons, and manifests as a winding exciton-photon coupling [see Eq.~\eqref{eq:winding-coupling}]. In a simplified picture one can think of the coupling of time reversal broken excitons to the photons as providing an effective magneto-optical effect that realizes Haldane and Raghu's proposal.

In the original proposal for topological polaritons \cite{karzig_topological_2015}, the mechanism that creates the topology is different and not given by a gapping out of Dirac cones [see Fig.~\ref{fig:comparison}(c)]. The topology rather emerges from a twist in the hybridized polariton bands that is created at the ring of resonance between excitons and photons when the exciton-photon coupling has a winding character. Since the topological gap is opened along the edge of the Brillouin zone [rather then at single points as in (a) and (b)] this proposal requires a more careful tuning in order to maximize the value of the topological gap. 

Although the topology providing mechanisms are different in case (b) and (c), they still rely on similar ingredients which are applied in different regimes: A single TE (or equivalently TM) photonic mode coupled to a single $J_z=\pm1$ excitonic mode. Completely separating the TE mode will in general require a periodic photon potential to create a band gap for the TM mode. Note, however, that the effect of the periodic photonic potential is more pronounced in proposals (a) and (b) because it is also crucial in creating the photonic Dirac cones. Splitting the two exciton species requires a strong enough magnetic field and the corresponding topological gap will always be limited by the strength of the exciton Zeeman field. Moreover, all known examples that exploit the exciton-photon resonance also require a periodic exciton potential to remove "accidental" degeneracies due to the essentially flat exciton dispersion. A promising approach without the need of a periodic exciton potential creates topology similar to proposal (c) by hybridizing two polariton bands\cite{nalitov_polariton_2015,bardyn_topological_2015} (rather then an exciton and a photon band). In this case the necessary ingredients are a periodic polariton potential (of either photonic or excitonic origin), and again a Zeeman field and a finite TE/TM splitting.

\section{Discussion}

We now discuss experimentally relevant parameters of the proposal and estimate the corresponding gap sizes. The main requirement is a photonic Dirac cone strongly coupled to a single excitonic band with similar energies. Realizing photonic Dirac cones is routinely done using photonic crystals \cite{joannopoulos_photonic_2011}, where a modulation of the dielectric constant acts similarly as the periodic potential discussed in this paper. The required lattice constant of the photonic crystal in order to create Dirac points at typical exciton energies (e.g. $\omega_0^X=1.6\mathrm{eV}$ in CdTe-based quantum wells) can be estimated by $c K \sim \omega_0^X$ and is of the order of $a_s\sim 200\mathrm{nm}$ (using $K=4\pi/3a_s$, and a dielectric constant of $\epsilon_r=8$). Note that larger values of $a_s$ could be realized when using smaller values of $\omega_0^X$ or Dirac points that form at higher bands in the Brillouin zone. Finally, the strong coupling between excitons and photons can be realized by forming a photonic cavity with an embedded exciton-supporting quantum well. The resulting couplings can be quite large, reaching $g_q=4\,\mathrm{meV}$ \cite{jiang_photonic_2014,karzig_topological_2015} already for a single quantum well.

The periodic exciton potentials can be externally induced by applying strain \cite{carusotto_quantum_2013} or surface acoustic waves to the quantum well \cite{de_lima_phonon-induced_2006,cerda-mendez_exciton-polariton_2013} leading to potential strengths of up to $u_X=1\mathrm{meV}$. Note that in order for the scheme to work the photonic and excitonic periodic potentials  should be commensurate. It might be possible to circumvent the corresponding fine tuning by internally creating a periodic exciton potential from exciton-exciton interactions \cite{bardyn_chiral_2015}. We imagine creating an exciton background by driving a mode of the photonic crystal (e.g. $q=0$) at energies far away from the photonic Dirac point. As an eigenmode, this exciton background will inherit the spatial profile of the photonic crystal. Finite exciton-exciton interactions then lead to an appropriate periodic potential for the excitons of interest (with energies close to the photonic Dirac cone). The corresponding interaction-induced blueshifts can reach values of up to 1meV \cite{amo_light_2010}.

Using Eq.~\eqref{eq:gap}, the above parameters allow for a topological gap of the order of $1 \mathrm{meV}$. Note that throughout this paper we only focused on a single ($J_z=+1$) exciton mode. The presence of the $J_z=-1$ mode (coupling to the photons with an opposite winding as in Eq.~\eqref{eq:winding-coupling}) then limits the size of the gap to the exciton-Zeeman splitting \cite{karzig_topological_2015} (for typical exciton g-factors of 2, a Zeeman energy of $1 \mathrm{meV}$ corresponds to magnetic fields of $8.5 \mathrm{T}$). 

In conclusion, we found that excitons in a magnetic field can be used to gap out photonic Dirac cones in a topological nontrivial way, leading to chiral polaritonic edge modes. The underlying mechanism can be understood in terms of the Haldane and Raghu proposal \cite{haldane_possible_2008,raghu_analogs_2008} where the required magneto-optical effect (here at optical frequencies) is induced by a resonant coupling to Zeeman-split excitons. This picture thus allows to connect the Haldane and Raghu proposal to previous works on topological polaritons that use similar ingredients in a different parameter regime. The latter allows for creating topological gaps at exciton-photon- \cite{karzig_topological_2015} or polariton-polariton resonances \cite{nalitov_polariton_2015,bardyn_topological_2015}. Interestingly, we showed that using photonic crystals with corresponding photonic Dirac cones can enhance the size of the topological gap for topological polaritons as compared to the original proposal of Ref.~\onlinecite{karzig_topological_2015}.

\begin{acknowledgments}
We thank Gil Refael and Charles-Edouard Bardyn for valuable discussions. This work was funded by the Institute for Quantum Information and Matter, an NSF Physics Frontiers Center with support of the Gordon and Betty Moore Foundation through Grant GBMF1250, and NSF through Grant DMR-1410435.
\end{acknowledgments}

\bibliography{topolaritons}	

\end{document}